\documentclass[conference]{IEEEtran}
\IEEEoverridecommandlockouts

\def\BibTeX{{\rm B\kern-.05em{\sc i\kern-.025em b}\kern-.08em
    T\kern-.1667em\lower.7ex\hbox{E}\kern-.125emX}}
    
{}
{}
{}

\usepackage{hyperref}
\usepackage{eurosym}

\usepackage{amsmath}
\usepackage{tikz}
\usepackage{mathdots}
\usepackage{yhmath}
\usepackage{cancel}
\usepackage{color}
\usepackage{siunitx}
\usepackage{array}
\usepackage{multirow}
\usepackage{amssymb}
\usepackage{gensymb}
\usepackage{tabularx}
\usepackage{booktabs}
\usetikzlibrary{fadings}
\usetikzlibrary{patterns}
\usetikzlibrary{shadows.blur}

\usetikzlibrary{quantikz}
\usepackage{adjustbox}

\usepackage{amsthm}
\usepackage{braket}

\begin{document}

\title{\text{\huge{Towards a Distributed Quantum Computing Ecosystem}}
\\ \small{(Invited Paper)}
}

\author{\IEEEauthorblockN{Daniele Cuomo}
\IEEEauthorblockA{\textit{University of Naples Federico II} \\
daniele.cuomo@unina.it}
\and
\IEEEauthorblockN{Marcello Caleffi}
\IEEEauthorblockA{\textit{University of Naples Federico II} \\
marcello.caleffi@unina.it}
\and
\IEEEauthorblockN{Angela Sara Cacciapuoti}
\IEEEauthorblockA{\textit{University of Naples Federico II} \\
angelasara.cacciapuoti@unina.it}
}

\maketitle

\begin{abstract}
The Quantum Internet, by enabling quantum communications among remote quantum nodes, is a network capable of supporting functionalities with no direct counterpart in the classical world. Indeed, with the network and communications functionalities provided by the Quantum Internet, remote quantum devices can communicate and cooperate for solving challenging computational tasks by adopting a distributed computing approach. The aim of this paper is to provide the reader with an overview about the main challenges and open problems arising with the design of a \textit{Distributed Quantum Computing ecosystem}. For this, we provide a survey, following a bottom-up approach, from a communications engineering perspective. We start by introducing the Quantum Internet as the fundamental underlying infrastructure of the Distributed Quantum Computing ecosystem. Then we go further, by elaborating on a high-level system abstraction of the Distributed Quantum Computing ecosystem. Such an abstraction is described through a set of logical layers. Thereby, we clarify dependencies among the aforementioned layers and, at the same time, a road-map emerges.
\end{abstract}

\maketitle

\section{Introduction}
\label{sec:1}

Nowadays, a tremendous amount of heterogeneous players entered the quantum race, ranging from tech giants - such as IBM and Google in fierce competition to build a commercial quantum computer - to states and governments, with massive public funds to be distributed over the next years \cite{CalCacBia-18,CacCalTaf-20,CacCalVan-20,CalChaCuo-20}.

In 2017, the European Commission launched a \euro{}1-billion flagship program to support the quantum research for ten years starting from 2018, and a first \euro{}132-million tranche is being provided during the following three years \cite{Car-18}. In 2018, the United States of America launched the National Quantum Initiative funded with \$1.2-billion over ten years and China is keeping up, investing billions to commercialize quantum technologies \cite{Gib-19}.

These huge efforts are justified by the disruptive potential of a quantum computer, beyond anything classical computers could ever achieve. Indeed, by exploiting the rules of quantum mechanics, a quantum computer can tackle classes of problems that choke conventional machines. These problems include chemical reaction simulations, optimization in manufacturing and supply chains, financial modeling, machine learning and enhanced security \cite{Bou-17, SchSinPet-15, GotLoLut-04}. Hence, the quantum computing has the potential to completely change markets and industries.

At the end of 2019, Google achieved the so-called \textit{quantum supremacy}\footnote{The term was coined by J. Preskill in 2012 \cite{Pre-12} to describe the moment when a programmable quantum device would solve a problem that cannot be solved by classical computers, regardless of the usefulness of the problem.} with a 54-qubits quantum processor, named \textit{Sycamore}, by sampling from the output distribution of 53-qubits random quantum circuits \cite{FraKunBab-19}. By neglecting some performance-enhancing techniques as pointed out by IBM \cite{PedGunMasGam-19, PedGunNan-19}, Google estimated that ``\textit{a state-of-the-art supercomputer would require approximately 10,000 years to perform the equivalent task}'' that required just 200 seconds on Sycamore.

By ignoring the noise effects and by coarsely oversimplifying, the computing power of a quantum computer scales exponentially with the number of qubits that can be embedded and interconnected within \cite{CacCalTaf-20}. And one of the reasons lays in a principle of quantum mechanics known as \textit{superposition principle}. Specifically, a classical bit encodes one of two mutually exclusive states - usually denoted as 0 and 1 - being in only one state at a certain time. Conversely, a qubit can be in an extra mode - called superposition – i.e., it can be in a combination of the two basic states \cite{CacCalTaf-20,CacCalVan-20}.

Hence -- thanks to the superposition principle -- a qubit offers  \textcolor{black}{richer} opportunities for carrying information and computing, since it can do more than just flipping between 0 and 1. And this quantum advantage grows exponentially with the number of qubits.
In fact, while $n$ classical bits are only in one of the $2^{n}$ possible states at any given moment, an $n$-qubit register can be in a superposition of all the $2^{n}$ possible states.

To give a flavor of the above, let us consider one of the killer applications of  \textcolor{black}{the} quantum computing: chemical reaction simulation \cite{Cao-19}. As highlighted in \cite{Dro-18}, the amount of information needed to fully describe the energy configurations of a relatively simple molecule such as caffeine is astoundingly large: $10^{48}$ bits. For comparison, the number of atoms in the Earth is estimated between $10^{49}$ and $10^{50}$ bits. Hence, describing the energy configuration of caffeine at one single instant needs roughly a number of bits comparable to $1$ to $10$ per cent of all the atoms on the planet. But this energy configuration description becomes suddenly feasible with a quantum processor embedding roughly $160$ \textit{noiseless} qubits, thanks to the superposition principle.

Unfortunately, qubits are very fragile and easily modified by interactions with the outside world, via a noise process known as \textit{decoherence} \cite{CalCac-19, CacCalVan-20}. Indeed, decoherence is not the only source of errors in quantum computing. Errors practically arise with any operation on a quantum state. However, isolating the qubits from the surrounding is not the solution, since the qubits must be manipulated to fulfill the communication and computing needs, such as reading/writing operations. Moreover, the challenges for controlling and preserving the quantum information embedded in a single qubit get harder as the number of qubits within a single device increases, due to coupling effects. In this regard, \textit{Quantum Error Correction} (QEC) represents a fundamental tool for protecting quantum information from noise and faulty operations  \textcolor{black}{\cite{NieChu-10, Bab-15}}.
 \textcolor{black}{However, QEC} operates by spreading the information of one \textit{logical} qubit into several physical qubits. Hence, solving problems of practical interest, such as integer factorization -- which constitutes one of the most widely adopted algorithms for securing communications over the  \textcolor{black}{current} Internet -- or molecule design may require millions of physical qubits \cite{Gib-19, Bou-17}.

Hence, on one hand researchers worldwide are leveraging on the advancement of different technologies for qubit implementation – superconducting circuits, ion traps, quantum dots, and diamond vacancies among the others – and innovative QEC techniques to scale the number of qubits beyond two-digits. On the other hand, the \textit{Quantum Internet}, i.e., a network enabling quantum communications among remote quantum nodes, has been recently proposed as the key strategy to significantly scale up the number of qubits \cite{Kim-08, Pir-16, CalCacBia-18, WehElkHan-18, Lon-19}.

 \textcolor{black}{In fact, the availability of such a network and the adoption of a distributed computing  paradigm allows  us  to  regard  the  Quantum Internet – jointly –  as a virtual quantum computer with a number of qubits that scales linearly with the number of interconnected devices.}
 
  \textcolor{black}{In this light, the aim of this paper is to provide the reader with an overview about the main challenges and open problems arising with the design of a Distributed Quantum Computing ecosystem}. 
 
  \textcolor{black}{We start in Sec.~\ref{sec:2} by introducing the Quantum Internet -- as the fundamental underlying communication infrastructure of a Distributed Quantum Computing ecosystem -- as well as some of its unique key applications. Then we go further in Sec.~\ref{sec:3}, by conceptualizing a high-level system abstraction of the Distributed Quantum Computing ecosystem from a communication engineering perspective. Such an abstraction is described through a set of (logical) layers, with the higher depending on the functionalities provided by the lower ones.  Thereby, we clarify dependencies among the aforementioned layers. Since, each layer of the ecosystem has some related open challenges, within Sec.~\ref{sec:4} we survey such challenges and open problems.}
 \textcolor{black}{Finally, in Sec.~{\ref{sec:5}} we conclude the paper with some perspectives.}

\section{The Quantum Internet}
\label{sec:2}

As mentioned  \textcolor{black}{in Sec.~\ref{sec:1}}, one promising approach to address the challenges arising with large-scale quantum processor   \textcolor{black}{realization} is to mimic modern high-performance computing  \textcolor{black}{infrastructures --} where thousands of processors, memories and storage units are inter-connected via a communication network, and the computational  \textcolor{black}{tasks} are solved by adopting a distributed approach.

 \textcolor{black}{To this aim}, it is mandatory to design and deploy the   \textcolor{black}{\textit{Quantum Internet}}, which formally defines a  \textcolor{black}{global} quantum network\footnote{ \textcolor{black}{We refer the reader to \cite{KozWehVan-20,ChoAkbRui-20} for a discussion about the differences underlying the notion of ``\textit{Quantum Internet}'' versus ``\textit{quantum network}''}.} able to transmit qubits  \textcolor{black}{and} to  \textcolor{black}{distribute entangled quantum states\footnote{The deepest difference between classical and quantum mechanics lays in the concept of \textit{quantum entanglement}, a sort of correlation with no counterpart in the classical world.  \textcolor{black}{For an in-depth introduction to} quantum entanglement we refer the reader to the classical book \cite{NieChu-10} \textcolor{black}{, whereas a concise description can be found in} \cite{CacCalVan-20}.} among} remote quantum devices  \textcolor{black}{\cite{CacCalVan-20, CalCacBia-18,CacCalTaf-20,Kim-08,Pir-16,WehElkHan-18,KozWehVan-20}}.

\begin{figure}[t]
    \centering
    \includegraphics[width=1\columnwidth]{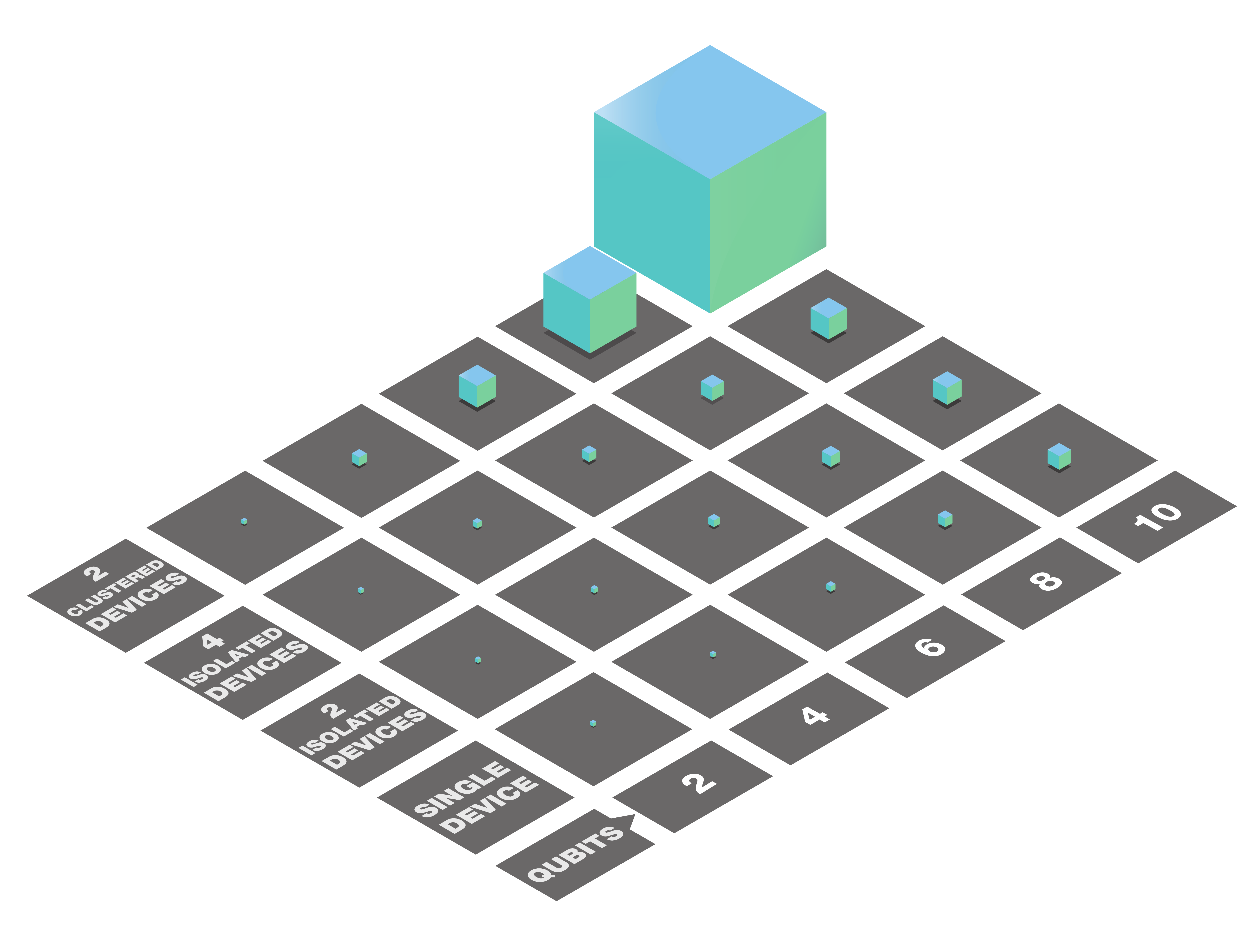}
    \caption{Distributed Quantum Computing speed-up. The volume of cubes represents the ideal  \textcolor{black}{quantum computing} power, i.e., in absence of noise and errors.  \textcolor{black}{As evident by comparing the power available at isolated devices versus the power achievable through \textit{clustered} devices, the interconnection of} quantum processors via the Quantum Internet provides an exponential  \textcolor{black}{computing} speed-up with respect to isolated devices.}
    \label{Fig:01}
\end{figure}

In fact,the availability of the corresponding  \textcolor{black}{underlying} network infrastructure and the adoption of  \textcolor{black}{the} distributed  \textcolor{black}{quantum} computing  \textcolor{black}{ paradigm} \cite{VanDev-16} allows  \textcolor{black}{us} to regard the Quantum Internet  \textcolor{black}{-- jointly -- } as a virtual quantum computer with a number of qubits that scales linearly with the number of interconnected devices. Hence,  \textcolor{black}{the Quantum Internet may enable} an exponential speed-up \cite{Yim-04,VanDev-16,CalCacBia-18} of the quantum computing power with just a linear amount of physical resources, represented by the  \textcolor{black}{interconnected} quantum processors. Indeed, by comparing the computing power achievable with quantum devices working independently  \textcolor{black}{versus} working as a unique quantum cluster, the gap comes out - as depicted in  \textcolor{black}{Fig.}~\ref{Fig:01} \cite{CalCacBia-18}.

Specifically, increasing the number of isolated devices lays to a linear speed-up, with a double growth in computational power by doubling the number of devices. Conversely, increasing the number of  \textcolor{black}{clustered} devices provides an exponential growth, with a significant advantage clearly visible with just two interconnected devices. For instance, a single $10$-qubit processor can represent $2^{10}$ states thanks to the superposition principle \textcolor{black}{, hence two isolated $10$-qubit processors can represent $2^{11}$ states.} But if we  \textcolor{black}{\textit{interconnect} the} two processors, the resulting virtual device can represent up to $2^{18}$ states \cite{CalCacBia-18}, depending on the number of qubits  \textcolor{black}{devoted to fulfill the communication needs of the clustered processors as discussed in Sec.~{\ref{sec:4.2}}.}

Before analyzing with further details in Sec.~\ref{sec:3} the  \textcolor{black}{resulting Distributed Quantum Computing ecosystem} from a communication engineering perspective, it is worthwhile to note that the availability of the Quantum Internet infrastructure enables unparalleled capabilities not restricted to the distributed computing \cite{WehElkHan-18,CacCalTaf-20}. Specifically, applications such as \textit{blind computing}, \textit{secure communications} and \textit{noiseless communications} have already been theorized or even experimentally verified \textcolor{black}{, as recently overviewed by an IETF Quantum Internet Draft~\cite{KozWehVan-20}}.

Blind quantum computing \cite{BroFitKas-09, AhaBenEba-10} refers to a server-client architecture where clients can send sensitive data to server, which elaborates inputs without knowing their values. This functionality allows to achieve a twofold goal: preserving data confidentiality as well as solving tasks that are intractable for the client -- that can be a classical computer -- but tractable for the server, which implements the quantum paradigm.

Secure communications in  \textcolor{black}{the} quantum field refers to the class of communication protocols that exploit quantum mechanics in order to get benefits unattainable using classical Internet. 
 \textcolor{black}{For instance, in the field of quantum cryptography, researchers study strategies for sharing keys among parties in total secrecy \cite{XuXiaLi-19, CheWanGan-19, SutOm-20}.} Whilst quantum byzantine agreement, a protocol used by  \textcolor{black}{multiple entities to distributively} agree on a common  \textcolor{black}{decision}, allows to achieve the  \textcolor{black}{consensus} in a constant number of rounds, whereas classical  \textcolor{black}{protocols scale} polynomially with the number of processors \cite{BenHas-05}.

Finally,  \textcolor{black}{the Quantum Internet provides the underlying infrastructure to achieve transmission rates exceeding the fundamental limits of conventional (quantum) Shannon theory \cite{CalCac-20,CacCal-19}.}  \textcolor{black}{Specifically, by} exploiting the capability of quantum particles to propagate simultaneously among multiple space-time trajectories,  \textcolor{black}{quantum} superpositions of noisy channels can behave as perfect  \textcolor{black}{noiseless} quantum communication channels, even if no quantum information can be sent throughout either of the  \textcolor{black}{noisy} component channels individually  \textcolor{black}{\cite{GyoImrNgu-18}}.

\section{Distributed Quantum Computing Ecosystem}
\label{sec:3}

The overall aim of classical distributed computing is to deal with hard computational problems by splitting out the computational tasks among several classical devices \textcolor{black}{,} in order to lighten the loads on single devices.

As mentioned in Sec.~\ref{sec:2}, with the network infrastructure provided by the Quantum Internet, this paradigm can be extended to quantum computing as well: remote quantum devices can communicate and cooperate for solving computational tasks by adopting a distributed computing approach  \textcolor{black}{\cite{Yim-04,VanDev-16}}. Since an entirely new paradigm -- characterized by unconventional phenomena ruled out by the quantum mechanics  \textcolor{black}{\cite{CalCacBia-18}, such as \textit{no-cloning} and \textit{entanglement}} -- is involved, a new  \textcolor{black}{\textit{ecosystem}} needs to be engineered.

\begin{figure}[t]
    \centering
    \includegraphics[scale=0.35]{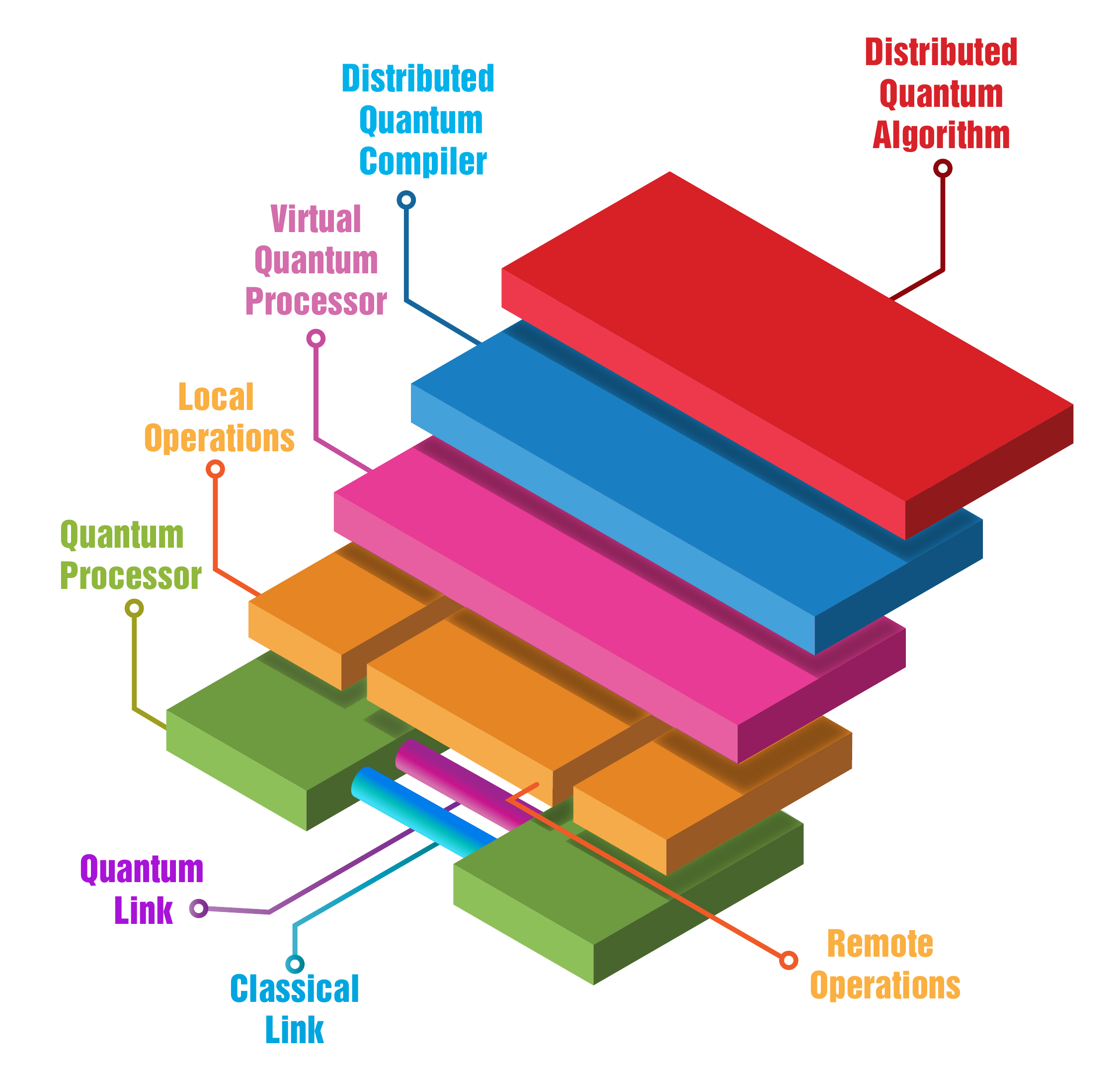}
    \caption{A high-level system abstraction of  \textcolor{black}{the} \textit{distributed quantum computing ecosystem}. The lowest  \textcolor{black}{layer} provides the communication/network functionalities and consists of quantum  \textcolor{black}{processors} interconnected with both classical and quantum links. Thanks to  \textcolor{black}{the} underlying communication infrastructure, both local and remote qubit operations can be executed. Hence, from a computing perspective, the two lowest levels concur to build a \textit{virtual quantum processor} with a number of qubits that scales with the number of  \textcolor{black}{inter-connected physical quantum processors. The virtual processor acts as an interface for the \textit{distributed quantum compiler}}, which maps  \textcolor{black}{a} quantum algorithm  \textcolor{black}{into} a sequence of local and remote operations \textcolor{black}{,} so that the available computing resources are optimized with respect to both the hardware and the network constraints.}
    \label{Fig:02}
\end{figure}

The infographic in  \textcolor{black}{Fig.}~\ref{Fig:02} is a stack depicting dependencies among a possible set of layers that together provide  \textcolor{black}{the \textit{Distributed Quantum Computing ecosystem}}. For the sake of clarity, we restrict our attention to  \textcolor{black}{a network} composed by two quantum  \textcolor{black}{processors}, directly inter-connected. Nevertheless, the discussion in the following can be easily extended to more complex  \textcolor{black}{network topologies}, provided that end-to-end routing and network functionalities such as  \textcolor{black}{\cite{Cal-17, GyoImr-18, PirDur-19, ImrGyo-18, ChaRoz-19, ShiQia-19} are available}.

Starting from bottom, in Fig.~\ref{Fig:02} we have the communication infrastructure underlying the Quantum Internet: spatially remote quantum devices able to communicate quantum information by means of a synergy of both classical and quantum communication  \textcolor{black}{resources}. Indeed, as  \textcolor{black}{we overview} in Sec.~\ref{sec:4.2}, the transmission of quantum information generally requires the transmission of classical information as well, hence the availability of a classical network infrastructure -- such as the classical Internet -- is required \cite{CacCalTaf-20,CacCalVan-20,WehElkHan-18}. This constraint has been highlighted by interconnecting  \textcolor{black}{the two quantum processors} with both a classical and a quantum link.

By exploiting the communication functionalities provided by the lowest level, both local and remote qubit operations can be executed. Specifically, the local operations -- i.e., operations between qubits  \textcolor{black}{stored within the same quantum processor} -- can be executed by exploiting the physical (or logical, whether the quantum device should natively implement QEC functionalities \cite{GamChoSte-17}) controls and readouts functionalities provided by the device. Conversely, the remote operations -- i.e., the operations between qubits  \textcolor{black}{stored at} different quantum devices -- pose further constraints, as discussed in Sec.~\ref{sec:4}.

Thanks to the abstraction provided by the two layers residing at the very bottom, a  \textcolor{black}{\textit{virtual quantum processor}} is obtained, where remote qubits are interconnected through virtual connections made possible by the remote operations. Clearly, remote operations are likely to be characterized by delays and error rates  \textcolor{black}{notably} larger than  \textcolor{black}{those characterizing} local operations. Hence, local operations should be preferred over remote ones as much as possible, even though remote operations are unavoidable whenever the number of qubits required to perform the computational task exceeds the number of qubits available  \textcolor{black}{at} a single device. Hence, an optimization must be performed by  \textcolor{black}{the \textit{distributed quantum compiler}}, so that the different operations required by the quantum algorithm as well as the input to the algorithm itself are  \textcolor{black}{properly} allocated among the qubits of the different devices.

Finally, at the very top we have the quantum algorithm, which is completely independent and unaware of the physical/logical constraints imposed by both the hardware and  \textcolor{black}{network particulars}, thanks to the abstraction provided by the underlying levels.  \textcolor{black}{We can think at this module as a minimal service for defining quantum algorithms, but also, in a wider perspective, as a platform where interesting functionalities are available - allowing, for instance, quantum machine learning \cite{TacMacGer-19} or quantum optimization algorithms \cite{VerArrBra-19, FarHar-16}}.

\begin{figure}[t]
    \centering
    \input{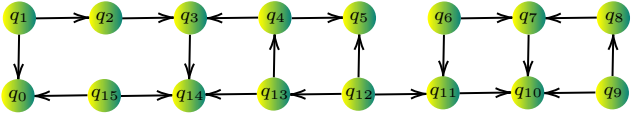}
    \caption{Coupling map of the  \textcolor{black}{\texttt{\textbf{ibmqx3}}} architecture: the nodes represent the qubits while the edges represent the possibility to have interactions between two qubits, i.e., to implement the \texttt{CNOT} operation.  \textcolor{black}{As an instance}, a \texttt{CNOT} operation can be directly executed between qubits $q_1$ and $q_2$ but not between  qubits $q_1$ and $q_3$ .}
    \label{Fig:03}
\end{figure}

\begin{figure}
        \begin{adjustbox}{width=\columnwidth}
    \begin{tikzcd}
        \lstick{$\ket{\psi_{q_1}}$} & \ctrl{1}\gategroup[2,steps=1, style={dashed, rounded corners, fill=green!20, inner xsep=2pt}, label style={label position=below, yshift=-0.4cm}, background]{\sc cnot} & \gate{H}\gategroup[2,steps=3, style={dashed, rounded corners, fill=yellow!20, inner xsep=2pt}, background]{\sc reversed cnot}
        & \ctrl{1} & \gate{H} & \ctrl{1} & \qw & \ctrl{1}\gategroup[2,steps=5,style={dashed,
rounded corners,fill=blue!20, inner xsep=2pt},
background]{{\sc swap}} & \gate{H} & \ctrl{1} & \gate{H} & \ctrl{1} & \qw \\
        \lstick{$\ket{\psi_{q_2}}$} & \targ{} & \gate{H}
        & \targ{} & \gate{H} & \targ{} & \ctrl{1} & \targ{} & \gate{H} & \targ{} & \gate{H} & \targ{} & \qw \\
        \lstick{$\ket{\psi_{q_3}}$} & \qw & \qw &\qw &\qw &\qw & \targ{} &\qw &\qw &\qw &\qw & \qw &\qw
    \end{tikzcd}
	\end{adjustbox}
    \caption{A \texttt{CNOT} operation between non-adjacent qubits can be implemented through a sequence of swapping operations, with each swap consisting of three \texttt{CNOT} (with the in-between \texttt{CNOT} being reverse, i.e., with target and control qubits swapped) between adjacent qubits \cite{ZulPalWil-19}.  \textcolor{black}{Thus, the circuit performs a $\texttt{CNOT}$ between $q_1$ and $q_3$, leaving $q_2$ unaltered. Note that $H$ denotes the Hadamard gate mapping a basis state into an even superposition of the basis states \cite{RiePol-11}, and that} the quantum state stored within qubit $q_i$ is denoted  \textcolor{black}{--} by adopting the standard bra-ket notation for describing quantum states  \textcolor{black}{--} as $\ket{\psi_{q_i}}$.}
    \label{Fig:04}
\end{figure}
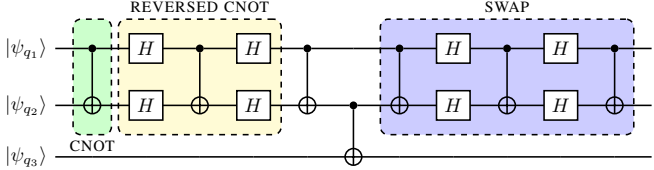

 \textcolor{black}{However,} several complications have been omitted so far. Indeed, communication protocols intrinsically imply overhead \textcolor{black}{, i.e.,} computing resources need to be dedicated, in order to deal with transmission processes and errors correction.  \textcolor{black}{Therefore, it's} worth going further towards this discussion, expanding components of the proposed  \textcolor{black}{ecosystem} and considering open challenges related with.

\section{Open Challenges Ahead}
\label{sec:4}
The aim of this section is  \textcolor{black}{to describe and to discuss some of the open problems related with the proposed Distributed Quantum Computing ecosystem. For this, some layers are individually discussed. Specifically, in Sec.~\ref{sec:4.1}} quantum processors are considered, paying particular attention to drawbacks induced by local operations involving multiple qubits.  \textcolor{black}{Sec.~\ref{sec:4.2}} introduces the interconnection between remote quantum processors, discussing the \textit{quantum teleportation} as a mean to transfer  \textcolor{black}{quantum} information between interconnected devices. After that, in  \textcolor{black}{Sec.~\ref{sec:4.3}} we discuss the gate teleportation as a strategy to perform remote operations.  \textcolor{black}{Sec.~\ref{sec:4.4} describes the layer responsible for abstracting and optimizing the execution of the quantum algorithms,} based on the characteristics of the underlying system.  \textcolor{black}{Finally, in Sec.~\ref{sec:4.4} we discuss some of the current standardization efforts.}

\subsection{Quantum  \textcolor{black}{Processor}}
\label{sec:4.1}
The most basic element of a distributed quantum computing ecosystem is identifiable with  \textcolor{black}{the local quantum computing device}. Here the qubits are connected  \textcolor{black}{according to} some directed and connected graph  \textcolor{black}{-- namely, the coupling map --} that accounts for the hardware limitations resulting from controlling and preserving the quantum information from decoherence and noise. As an example,  \textcolor{black}{Fig.~\ref{Fig:03} depicts the coupling map of the \texttt{\textbf{ibmqx3}}} quantum processor  \cite{ZulPalWil-19}, with nodes representing qubits while edges represent the possibility to have interactions between two qubits, i.e., to implement one of the fundamental quantum operations: the \texttt{CNOT} operation\footnote{The \texttt{CNOT} operation involves two qubits, referred to as control qubit and target qubit. It works as follows \cite{RiePol-11}: if the control qubit is 1, then the target qubit value is flipped. Otherwise, nothing happens.}.  \textcolor{black}{In fact}, there exists  \textcolor{black}{a universal} quantum gate set\footnote{ \textcolor{black}{A universal} gate set is any set of gates which any operation possible on a quantum computer can be reduced to \cite{Ray-15}.} with \texttt{CNOT} as the only operator belonging the set that involves more than one qubit. Thus, we can focus on problems related to the \texttt{CNOT} operation keeping the discourse general.

\begin{figure}[t]
    \centering
    \includegraphics[width=1\columnwidth]{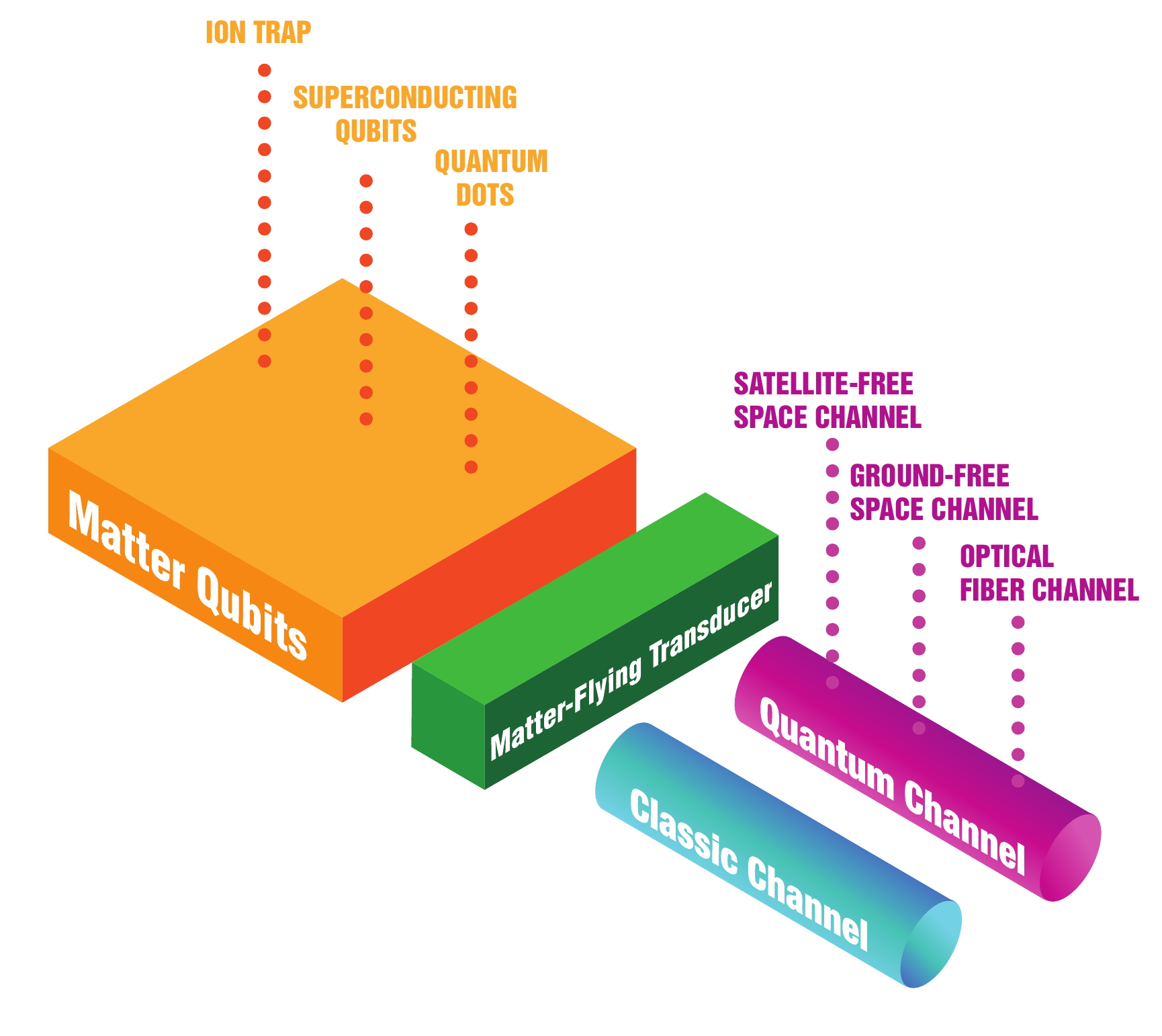}
    \caption{Pictorial representation of a matter-flying transducer, which is needed to convert matter qubits -- i.e., qubits for information processing/storing -- into flying qubits -- i.e., qubits for information transmission -- and vice versa.}
    \label{Fig:05}
\end{figure}

It is immediate to observe that only  \textcolor{black}{nodes -- i.e., qubits -- } linked by an edge can directly interact \cite{LinMasRoe-17}. Nevertheless, for an algorithm designer it is  \textcolor{black}{crucial} being able to define a circuit without restrictions on interactions. Indeed, a circuit programming model can easily abstract from this restriction \cite{NieChu-10}, resulting in a fully connected graph at the cost of an overhead due to indirect execution of the desired operation.  \textcolor{black}{For instance}, let us consider to carry out a \texttt{CNOT} between the two non-adjacent qubits $q_1$ and $q_3$ of  \textcolor{black}{Fig.}~\ref{Fig:03}. By performing two  \textcolor{black}{state swapping operations} for each edge belonging to the shortest-path from node $q_1$ towards node $q_3$, the overall result will be equivalent to a \texttt{CNOT} between $q_1$ and $q_3$, keeping other states unchanged.  \textcolor{black}{Fig.}~\ref{Fig:04} clarifies the process above.

The overhead induced by the swapping operations explains how important is the topological organization of devices and the circuit design as well.

\subsection{Quantum Link}
\label{sec:4.2}

As mentioned in Sec.~\ref{sec:3}, the  \textcolor{black}{Quantum Internet} requires both classical and quantum links. In this perspective, a distinction between \textit{matter} and \textit{flying} qubits -- i.e., between qubits for information processing/storing and qubits for information transmission -- must be made  \textcolor{black}{\cite{CacCalTaf-20}}.

As regards to the matter qubits, several candidate technologies are available, each one with its pros and cons \cite{VanDev-16}. Conversely, as regards to the flying qubits, there exists a general consensus about the adoption of photons as qubit substrate \cite{CacCalTaf-20}. However, heterogeneity arises by considering the different physical channels the photons propagate through, ranging from free-space optical channels (either ground or satellite free-space) to optical fibers.  \textcolor{black}{Thus, a transducer} for matter-flying conversion is needed as depicted in  \textcolor{black}{Fig.}~\ref{Fig:05} and discussed with further details in \cite{CacCalTaf-20}. And communication models need to take into account such  \textcolor{black}{a} technological heterogeneity \textcolor{black}{,} with the aim of providing a black box for upper protocol layers with one common logic. 

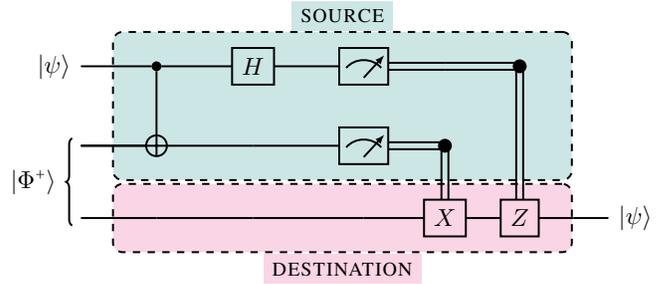
\begin{figure}
    \centering
    \begin{adjustbox}{width=1\columnwidth}
		\begin{tikzcd}
			\\
			\lstick{$\ket{\psi}$} & [12pt]\ctrl{1}  \gategroup[wires=2,steps=5,style={dashed,rounded corners,inner xsep=10pt,inner ysep=3pt, fill=teal!20}, background, label style={label position=above,  yshift=-0.0cm, fill=teal!20}, background]{\sc source}
				& [12pt]\gate[style={fill=teal!20}]{H} & [12pt]\meter[style={fill=teal!20}]{} & \cw & \cwbend{2} \\
			\lstick[wires=3]{$\ket{\Phi^\texttt{+}}$} & \targ & \qw & \qw & \meter[style={fill=teal!20}]{} & \cwbend{1} & \\
			& \gategroup[wires=1,steps=5,style={dashed,rounded corners,inner xsep=10pt,inner ysep=3pt, fill=magenta!20}, background,label style={label position=below, yshift=-0.44cm, fill=magenta!20}, background]{\sc destination}
				\qw & \qw & \qw & \gate[style={fill=magenta!20}]{X} & \gate[style={fill=magenta!20}]{Z} & \qw & \qw \rstick{$\ket{\psi}$} &
		\end{tikzcd}
	\end{adjustbox}
    \caption{Quantum teleportation circuit. First two wires belong to source, whereas the bottom wire belongs to destination.  \textcolor{black}{A generic state} $\ket{\psi}$ is initially stored at the source \textcolor{black}{. Once the teleportation process} is fulfilled, the original state is  \textcolor{black}{available} at the destination, regardless of its value. $\ket{\Phi^{\texttt{+}}}$ represents an EPR pair, that is a couple of maximally entangled qubits \cite{EinPodRos-35}. The result of the measurement process at the source is sent to the destination via a classical link \textcolor{black}{. The carried classical bits are thus used for determining whether gates $X$ and $Z$ -- corresponding to a bit- and a phase-flip, respectively \cite{RiePol-11} -- must be performed or not, in order to recover the original state $\ket{\psi}$ from the EPR pair member available at the destination.}}
	\label{Fig:06}
\end{figure}

 \textcolor{black}{Furthermore, quantum mechanics} does not allow an unknown qubit to be copied or  \textcolor{black}{even simply}  observed/measured.  \textcolor{black}{As a consequence, the} communication techniques utilized to interconnect spatially remote quantum devices cannot be directly borrowed from classical communications. In this context, \textit{quantum teleportation} is widely accepted as one of the most promising quantum communication technique between  \textcolor{black}{remote} quantum nodes \cite{BenBraCre-93, Van-12, CacCalVan-20}. Quantum teleportation has been experimentally verified \cite{RenXuYon-17} and it requires, as depicted in Fig.~\ref{Fig:06}, a pair of parallel resources\footnote{ \textcolor{black}{For an in-depth} discussion about the quantum teleportation process \textcolor{black}{,} we refer the readers to \cite{CacCalVan-20}.}. One of these resources is classical: two bits must be transmitted from the source to the destination. The other resource is quantum: an entangled pair of qubits must be generated and shared between the source and the destination. 

In the context of the distributed quantum computing ecosystem, quantum teleportation constitutes the foundation of a communication paradigm known as \textit{teledata} \cite{VanNemMun-06}, which generalizes the concept of moving state among qubits to remote devices.

To provide a concrete example of the teledata concept, a further distinction must be made between \textit{communication qubits} and \textit{data qubits}. Specifically, within each quantum device, a subset of  \textcolor{black}{matter qubits is reserved for fulfilling the communication needs to generate entanglement. These qubits are called} communication qubits  \textcolor{black}{\cite{KozWehVan-20}}, to distinguish them from  \textcolor{black}{the remaining matter qubits within the device devoted to processing/storage,} referred to as data qubits.

\begin{figure}[t]
    \centering
    \input{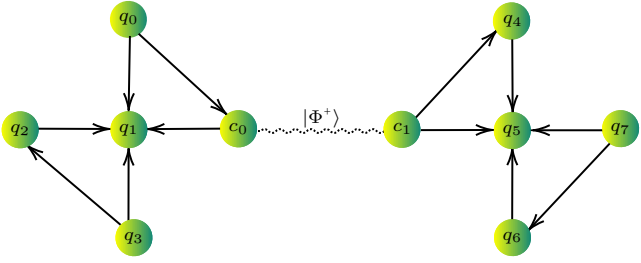}
    \caption{Coupling map representing physical architecture of two quantum devices  \textcolor{black}{\texttt{\textbf{ibmqx2}}} interconnected via  \textcolor{black}{the quantum teleportation paradigm}. Nodes labeled with $c_0$ and $c_1$  \textcolor{black}{denote \textit{communication qubits}}, and the dotted line indicates  \textcolor{black}{that they are in the entangled state $\ket{\Phi^{\texttt{+}}}$, i.e., they form an EPR pair}. Conversely, $q_i$ with $i \in \{0,1,2,3\}$   \textcolor{black}{denote \textit{data qubits} -- i.e., qubits available for computing tasks}.}
    \label{Fig:07}
\end{figure}

 \textcolor{black}{As an example}, consider two  \textcolor{black}{\texttt{\textbf{ibmqx2}} architectures \cite{ibmqx}} interconnected via quantum teleportation as depicted in  \textcolor{black}{Fig.}~\ref{Fig:07}.  \textcolor{black}{The $c_0$, $c_1$ pair is in the state $\ket{\Phi^{\texttt{+}}}$ - that is the standard notation to denote a couple of maximally entangled qubits \cite{NieChu-10}, also known as EPR pair \cite{EinPodRos-35}. Any} kind of interaction between remote devices involves communication qubits, but not all the data qubits are connected with them. As already explained in  \textcolor{black}{Sec.}~\ref{sec:4.1}, interactions between non-adjacent qubits are feasible but they imply an overhead. A solution would be adding more qubits to the communication set, but it means sacrifice further valuable resources  \textcolor{black}{for processing and storage}. For this reason,  \textcolor{black}{the selection of the communication qubit set is a crucial task within the distributed quantum computing ecosystem and it implies a carefully evaluation of the} trade-off between the number of data and communication qubits.

Next section shows how to exploit teleportation in order to not only send information but also perform joint operations between remote qubits.

\subsection{Teleporting Gates}
\label{sec:4.3}
Distributed quantum computation requires the capability to perform quantum operations on qubits belonging to remote quantum devices.

As mentioned in the previous section, one possible solution is to resort to the \textit{teledata} concept, by moving the quantum information from a quantum device to another via the teleportation process, through an entangled pair.

However, an entangled pair allows one to implement also a so-called teleporting gate, or \textit{telegate} \cite{Van-06}. From a theoretical perspective, we have already observed that providing the \texttt{CNOT} operation - together with other single qubit gates - is enough to perform any kind of quantum algorithm. Therefore, returning to stack dependencies of  \textcolor{black}{Fig.}~\ref{Fig:02}, we can conceptualize a service that provides a set of  \textcolor{black}{remote operations} based on teleported gates. Such service will directly interact with the physical system, exploiting the entanglement generation and distribution functionality \cite{CacCalVan-20}.

Specifically, by considering the topology depicted in  \textcolor{black}{Fig.}~\ref{Fig:07}, it is possible to implement the \texttt{CNOT} as the joint operation between qubits belonging to spatially remote devices. Indeed,  \textcolor{black}{by exploiting} the availability of two communication qubits -- $c_0$ and $c_1$ -- shared between the two remote devices and storing an EPR pair, it is possible to perform a remote \texttt{CNOT} operation\footnote{ \textcolor{black}{For a more comprehensive presentation of the telegate, we refer the reader to \cite{ChoBluWan-18}.}} with, for example, $q_0$ as control qubit and $q_4$ as target qubit.  \textcolor{black}{Fig.}~\ref{Fig:08} shows the corresponding circuit, consisting of local \texttt{CNOT} operations and single-qubit operations and measurements.

\subsection{Distributed Quantum Compiler}
\label{sec:4.4}

 \textcolor{black}{Concepts discussed so far provide} some fundamental underlying communication functionalities enabling the distributed quantum computing paradigm.

\begin{figure}[t]
    \centering
    \begin{adjustbox}{width=1\columnwidth}
		\begin{tikzcd}
			\lstick{$\ket{\psi}$} & \qw & \ctrl{1}\gategroup[wires=2,steps=7,style={dashed,rounded corners,inner xsep=10pt,inner ysep=3pt, fill=teal!20}, background, label style={label position=above,  yshift=-0.0cm, fill=teal!20}, background]{\sc source} & \qw & \qw & \qw & \qw & \qw & \gate[style={fill=teal!20}]{Z} & \qw & \qw \\
			\lstick[wires=2]{$\ket{\Phi^\texttt{+}}$} & \qw & \targ{} & \qw & \qw & \qw & \meter[style={fill=teal!20}]{} & \cwbend{2} \\
			\lstick{} & \qw & \ctrl{1}\gategroup[wires=2,steps=7,style={dashed,rounded corners,inner xsep=10pt,inner ysep=3pt, fill=magenta!20}, background,label style={label position=below, yshift=-0.44cm, fill=magenta!20}, background]{\sc destination} & \qw & \gate[style={fill=magenta!20}]{H} & \qw & \meter[style={fill=magenta!20}]{} & \cw & \cwbend{-2} \\
			\lstick{$\ket{\phi}$} & \qw & \targ{} & \qw & \qw & \qw & \qw & \gate[style={fill=magenta!20}]{X} & \qw & \qw & \qw
		\end{tikzcd}
	\end{adjustbox}
    \caption{ \textcolor{black}{Quantum telegate circuit implementing a} \texttt{CNOT} operation between remote qubits.  \textcolor{black}{Specifically, a \texttt{CNOT} operation between qubits placed at different devices -- say qubits $q_0$ and $q_4$ in Fig.~\ref{Fig:07}-- is performed through \texttt{CNOT}s between each qubit and the member of an EPR pair stored at the same device, followed by single-qubit gates and measurements. Note that $\ket{\psi}$ and $\ket{\phi}$ denote the generic initial states stored at $q_0$ and $q_4$, respectively, whereas $\ket{\Phi^{\texttt{+}}}$ denotes the EPR pair stored by the communications qubits -- say qubits $c_0$ and $c_1$ in Fig.~\ref{Fig:07}.}}
    \label{Fig:08}
\end{figure}
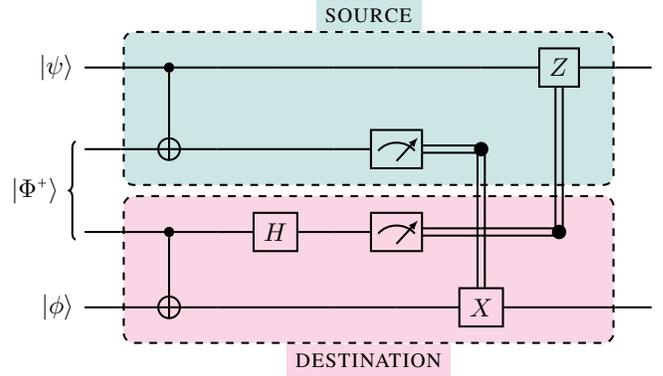

Indeed, the definition of a quantum algorithm can be very abstract. In general, it goes through the definition of a quantum circuit -- where a computation is a sequence of quantum gates on  \textcolor{black}{a register} -- as those  \textcolor{black}{depicted in Figs.~\ref{Fig:06} and \ref{Fig:08}. And algorithm designers} may benefit from an abstraction which hides the complications due to physical features.

However, as mentioned in Sec.~\ref{sec:3}, within the  \textcolor{black}{Virtual Quantum Processor} remote qubits belonging to remote quantum devices are interconnected through virtual connections made possible by the remote operations as described in Secs.~\ref{sec:4.2} and \ref{sec:4.3}. Unfortunately, remote operations are likely to be characterized by delays and error rates larger than local operations.  \textcolor{black}{The reason underlying this statement is that each protocol step needed for realizing remote operations -- from the entanglement distribution to the gate operations -- is affected by decoherence \cite{CacCalVan-20}, i.e., by a quantum-specific noise process. Decoherence affects local operations as well. However, since the effects of such a noise become stronger as function of the time \cite{CacCalVan-20,CacCalTaf-20}, remote operations, by involving further away parties, are more vulnerable.}  \textcolor{black}{It follows that, given a particular quantum circuit describing a quantum algorithm, the Distributed Quantum Compiler is required to optimize the circuit}\footnote{ \textcolor{black}{Note that a circuit} optimization is needed also in case of a single quantum processor \cite{ZulPalWil-19, FerAmo-18}  \textcolor{black}{to reduce the overhead arising with the swap operations discussed in Sec.~\ref{sec:4.1}}.} so that the number of remote operations is minimized as much as possible  \textcolor{black}{to limit the decoherence effects}.

Furthermore, the compiler should be able to optimize the circuit so that it can be executed regardless of the underlying network topology. Indeed, it may be the case that two remote quantum devices are not directly connected -- i.e., no  \textcolor{black}{communication qubits} are shared between the two devices -- even though some remote operations between qubits stored at these devices are required. Hence, the compiler should optimize the corresponding quantum circuit so that the entanglement swapping operations\footnote{ \textcolor{black}{Entanglement swapping is a technique used to entangle distant nodes -- without physically sending an entangled qubit through the entire distance -- by \textit{swapping} the entanglement generated at intermediate nodes \cite{Van-12}}.} among remote devices are minimized. Finally, as discussed in Sec.~\ref{sec:4.2}, there exists a trade-off between data and communication qubits. The larger is the number of communication qubits in a device, the higher is the rate of remote operations achievable at the price of reducing the number of qubits devoted to computation.


\subsection{What's Next}
\label{sec:4.5}
Considering modern available technologies, it seems reasonable to envision that we will see a first attempt of interconnection among quantum computers located nearby. Likely, one of the main tech companies currently providing cloud access to isolated quantum computers -- such as IBM \cite{ibmq}, HIQ \cite{hiq}, Amazon Braket \cite{aws} or Azure Quantum \cite{azure} -- will scale the available quantum computing power by interconnecting few quantum computers located few meters away within a \textit{quantum farm} \cite{CacCalTaf-20,CalChaCuo-20}. After that, consider that some of these companies will have quantum computers distributed over the world. Thus, it would not be surprising to see  interconnection among quantum computers or even among quantum clusters miles away from each other.

However, interconnecting quantum devices implies the need for a communication standard. In other words, remote devices -- likely to have technological differences -- will need to agree on network protocols in order to exchange interpretable information. Thus, the Quantum Internet will need a logical architecture as well as the classical Internet does -- i.e., with the TCP/IP Internet protocol suite being the standard \textit{de-facto}. Research in this direction has already started, within the Internet Engineering Task Force (IETF), where researchers are trying to conceptualize the Quantum Internet as a service-oriented platform \cite{ChoAkbRui-20, KozWehVan-20}.

\section{Conclusions}
\label{sec:5}
With this paper, we have presented a layered  \textcolor{black}{ecosystem that outlines a possible key strategy towards large-scale quantum processor design based on the distributed quantum computing paradigm. Within the envisioned ecosystem,} the lowest layers integrate the Quantum Internet as the fundamental underlying infrastructure providing networking and communication functionalities among remote quantum devices. Conversely, the upper layers are responsible for mapping the quantum algorithm onto the underlying physical infrastructure by optimizing the available computational resources as well as by accounting for the  \textcolor{black}{constraints} induced by the hardware/network configuration.

\bibliographystyle{IEEEtran.bst}
\bibliography{main.bbl}

\end{document}